\documentclass{mem}
\usepackage{natbib}
\usepackage{txfonts}
\usepackage{balance}
\usepackage{graphicx}
\usepackage{lmodern}
\usepackage[a4paper,breaklinks,dvipdfm]{hyperref}
\idline{75}{282}
\begin{document}
\def\teff{$T\rm_{eff }$}
\def\kms{$\mathrm {km s}^{-1}$}

\title{
Globular cluster-massive black hole interactions in galactic centers
}

   \subtitle{}

\author{
R. \,Capuzzo-Dolcetta
          }

\institute{
Dep. of Physics, Sapienza, universit\'a di Roma, 
piazzale A. Moro 2,
I-00185 Roma, Italy
\email{roberto.capuzzodolcetta@uniroma1.it}
}

\authorrunning{Capuzzo-Dolcetta}

\titlerunning{GCs and MBHs}

\abstract{
Many, if not all, galaxies host massive compact objects at their centers.
They are present as singularities (super massive black holes) or high density star clusters (nuclear tar clusters). In some cases they coexist, and interact more or less strongly. In this short paper I will talk of the {\it merger} globular cluster scenario, which has been shown in the past to be an explanation of the substantial mass accumulation in galactic centers. In particular, I will present the many astrophysical implications of such scenario pointing the attention on the mutual feedback of orbitally decaying globular clusters with massive and super massive black holes. 
}
\maketitle{}

\section{Introduction}

Galaxies of all the Hubble types have peculiarities in their inner region.
It is clear nowadays that brighter galaxies host massive and even super massive black holes (SMBHs) oscillating around their gravity centers while fainter galaxies often host very dense stellar aggregates, the so called nuclear star clusters (NSCs). The coexistence of an SMBH and a surrounding NSC is not uncommon, anyway.
In a recent work \citet{Geo16} showed that for stellar host masses above $\sim 5\times 10^{10}$ M$\odot$ the BH mass begins to dominate over the NSC mass, while for lower galaxy masses, the NSC outweighs the MBH.

A collection of recent data indicates that NSCs are present in 75\%  
of late-type spirals (Scd-Sm), 50\% of earlier type sp. (Sa-Sc), and 70\% of spheroidal (E and S0) galaxies.  
A relevant issue is that NSCs contain both a old ($\geq 1$ Gyr) and a young ($\leq 100$ Myr) stellar population. This gives an important constraint to NSC formation hypotheses. 
The presence in different galaxies of compact massive objects (CMOs), although different in structure and type, suggests the existence of some correlation between the galactic environment and the CMO, which requires a convincing theoretical interpretation. 

So far, two main frameworks have been suggested. One, called {\it in situ} model, is still more a hypothesis than a detailed model. It claims a local, central, star formation giving rise to a dense star cluster. Neither the origin of the gas (funnelled toward the centre by some angular momentum loss?) nor the modes of the necessarily efficient central star formation in presence of a massive black hole therein with its strong tidal disturbance effect, have been explained in a convincing way, so far. 
The other proposed explanation for galactic nucleus formation is via orbital decay and subsequent merger in the galactic center of massive globular clusters
This scheme was first suggested by \citet{Tre75} and later developed and made more straightforward by \citet{CD93} and a series of following papers of his research group. \\
It is, indeed, relevant to acknowledge that this infall and merging scheme, upon which many people have been working so far, has had the \citet{Tre75} and the \citet{CD93} as seminal papers. Actually, these two papers were able to quantify the possibility of carrying many stars in a compact structure around a galactic center forming a resolved stellar nucleus therein even before that what have been later called nuclear star clusters were discovered.

\section{Compact Massive Objects}

As said above, galaxies use to host massive objects in their central region.
The compactness of these CMOs is an increasing function of the parent galaxy luminosity (mass).
In Table \ref{scales} we give values of characteristic physical scales for some CMOs.

\begin{table*}
\caption{CMO characteristic parameters}
\label{scales}
\begin{center}
\begin{tabular}{ccccc}
\hline
\\
CMO & Mass (M$_\odot$) & Length (pc) & density (M$_\odot$ pc$^{-3}$)& location \\
\hline
\\
SMBH  &$10^6-10^{10}$ & $10^{-7}-10^{-2} $& $10^{27} $ &$\rm{gal.  ~center}$ \\
NSC  &$10^6-10^8$ & $4 $& $10^6 $ &$\rm {inner~ pc}s$ \\
GC  &$ 10^4-10^6$ & $2-5$& $10^3-10^6$ &$\rm {kpc}$ \\
\hline
\end{tabular}
\end{center}

\end{table*}

The values reported in the table essentially say that in order to make the infall and merger hypothesis viable, a shrink of the GC distribution length scale (i.e. a shrink of the spatial distribution of a sub sample of GCs) for a factor $1000$, reducing the kpc GC spatial distribution scale to the inner pc scale, is needed.
 
Is this possible? And how?

A straightforward positive answer to these, and other, questions related to the suitability of the merger hypothesis has already been done in many papers since \cite{CD93}.\\
Here I limit to give a qualitative physical insight to this topic.

GCs move as internally structured test \lq particles\rq in the external galactic potential.
Along the motion, different degrees of freedom are diversely excited: the interaction with the galactic field induces some reduction of the GC orbital energy
(quenching of what I call {\it external}, orbital degrees of freedom, that of the GC as a whole) and a contemporary excitation of {\it internal} degrees of freedom, i.e. a \lq heating\rq~ of the GC. The first phenomenon corresponds to dynamical friction, tending to shrink the GC distribution length scale in a rate directly proportional to the GC mass, and is strongly dependent upon the GC orbital distribution. The other phenomenon acts oppositely, in the sense that the larger the heating, the higher the evaporation of stars from the cluster with consequent reduction of its mass and thus of the efficiency of dynamical friction breaking with a following fate of GC dissolution in the field before it reaches an orbital equilibrium around the galactic center. The likelihood of a significant excitation of the internal degrees of freedom is, roughly, evaluated by this simple back of the envelope calculation.

Let $E$ be the GC binding energy per unit mass and $E_o$ its orbital energy per unit mass. Assuming that the GC, considered as a homogeneous sphere of mass $M_{GC}$ and size $R_{GC}$, moves in a homogeneous galactic bulge whose mass and size are $M_{b}$ and $R_{b}$, we have 
 
\begin{equation}
\frac{E}{E_o}= \frac{3}{5}\frac{M_{GC}}{M_{b}}\frac{R_{GC}}{R_{b}}
\left[\frac{3}{2}-\left(\frac{r}{R_{b}}\right)^2\right],
\end{equation}

which, assuming for both $M_{GC}/M_{b}=0.01$ and $M_{GC}/M_{b}=0.01$  as typical values for the GC moving  on a circular orbit of radius $r$ within the bulge ($r/R_b < 1$), leads to

\begin{equation}
\frac{E}{E_o}  \leq 9\times 10^{-5}.
\end{equation}

This means that a transfer of just $0.01 \%$ of the orbital energy into GC internal energy (i.e. into internal degrees of freedom) suffices to disperse a loose GC during its infall to the galactic center. GCs can be fragile systems.
Of course, the effectiveness of energy swap from the orbital \lq reservoir\rq~ to the external and internal GC degrees of freedom is a strongly non linear process that, to be thoroughly quantified, needs sophisticated $N$-body simulations of the GC motion in a particle-sampled galaxy environment.
We refer to \cite{AS16} for such work.

\section{Consequences of GC infall}

The GC infall and merger scenario is a framework which has many intriguing fallouts. We can list: i) NSC formation; ii) NSC vs host galaxy properties scaling relations; iii) mutual feedback with a local MBH (see Fig. 1); iv) high- and hyper-velocity star generation; v) solution of the inner pc problem?\\
I limit here to give some, non exhaustive, example references for the various points and to make, in the following subsection, an easy treatment of one specific point in topic ii) (scaling relations).

Some references:

\begin{itemize}
\item Point i): \cite{Tre75}; \cite{CD93}; \cite{Ant12}; \cite{Ant13}.
\item Point ii): \cite{Erw12}, \cite{Geo16}, \cite{TeM16}, \cite{CD16}. 
\item Point iii): \cite{CD93}, \cite{AS16}.
\item Point iv): \cite{CD15}, \cite{Fra16}.
\item Point v): no papers published, yet.
\end{itemize}

\begin{figure}[]
\includegraphics[width=65mm]{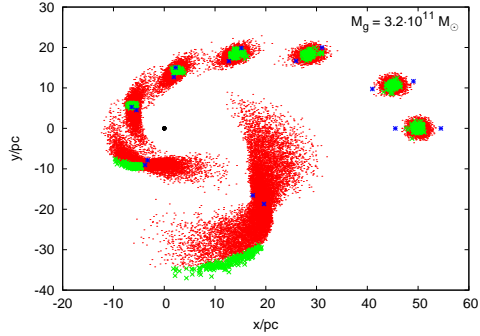}
\caption{
\footnotesize
Various snapshots of the GC moving in a counter clockwise motion on an eccentric orbit in a $3.2\times 10^{11}$ M$_\odot$ galaxy. Escaping stars are in green, while red dots identify the stars that remain bound to the cluster. The black filled circle labels the $5\times 10^8$ M$_\odot$ SMBH, while the blue asterisks represent the lagrangian points L1 and L2 (from \cite{AS16}.)
}
\label{fig1}
\end{figure}

\subsection{A straightforward correlation}

One fundamental correlation in the context of CMO studies is that linking the CMO mass and the velocity dispersion of the parent galaxy. While the SMBH mass vs galaxy velocity dispersion is a steep increasing function ($M_S \propto \sigma^5$, \cite{grah11}), the NSC mass seems to correlate to $\sigma$ with a shallower profile ($M_N \propto \sigma^{1.6}$, \cite{grah12}). Intriguingly, this shallower profile has a straightforward interpretation in the infall and merger scenario for NSC formation. 

This result can be derived, again, from a simple formal development. 
Following the derivation in \cite{AS14}, based on the assumption of GCs of equal mass $M$, spatially distributed according to a mass density power law  $\rho(r) \propto r^\alpha$ in a singular isothermal spherical galaxy ($\rho_g(r) \propto r^{-2}$) with mass $M_g$, (constant) velocity dispersion $\sigma$ and spatially cut at $R$, the nucleus mass resulting from GC merger is, at every time $t$ 
\begin{equation}
M_{\rm n} = f\frac{2}{G} {\left(0.6047G\ln \Lambda M\right)} ^{\alpha+3}t^{\frac{\alpha +3}{2}}
\frac{\sigma^{\frac{1-\alpha}{2}}}{R^{\alpha+2}},
\end{equation}
for $t\leq \sigma R^2_0/(0.6047G\ln \Lambda M)$, while  $M_{\rm n}(t)$ saturates to $M_{\rm GCS}$ at $t=\sigma R^2_0/(0.6047G\ln \Lambda M)$. 
\\
Equation 3 (in which $f$ is the fraction of the total GC mass to the galactic mass) is obtained by a straightforward analytical integration of the 1st order differential equation governing the orbital angular momentum evolution of the GC in the host galaxy. 
Note that eq. 3 reduces to the $M_{\rm n}-\sigma$ scaling relation, $M_{\rm n} \propto \sigma^{3/2}$, obtained by \cite{Tre75} in the case of $\alpha=-2$, i.e. for GC distributed the same way as the galactic isothermal background and is independent of the galactic radius $R$. This is the only case where the dependence on the galactic radius $R$ cancels out.
For other values of $\alpha$ in the allowed range, the dependence of $M_{\rm n}$ on $\sigma$, in the assumption of a virial relation between galactic $R$ and $M_{\rm g}$ ($R \propto M_{\rm g}/\sigma^2$), becomes
\begin{equation}
M_{\rm n}(t) \propto \frac{\sigma^\frac{9+3\alpha}{2}}{M_{\rm g}},
\end{equation}
which corresponds to a slope in the range from $0$ of the steeper ($\alpha = -3$)  GCS radial distribution to $9/2$ of the flat ($\alpha=0$) distribution.
\\
The relevant result here is that the slope of the $M_{\rm n}-\sigma$ relation in the regime of dynamical friction dominated infall process is expected to have an upper bound in any case smaller than that of the $M_{\rm BH} - \sigma$ relation.

\section{Conclusions}

The globular cluster infall and merging scenario is an extensively studied frame which represents an attractive self-consistent explanation of various astrophysical phenomena.
This approach was originally proposed by \cite{Tre75}
and later developed by \cite{CD93}. Many other authors have later followed the lines shown in those two seminal paper.
As a not exhaustive list of the topics connected to the infall and merger scenario, we indicate: 
\\ i) NSC formation; ii) NSC vs host galaxy properties scaling relations; iii) mutual feedback with a local MBH; iv) high- and hyper-velocity star generation;  v) solution of the inner pc problem.
 

\bibliographystyle{aa}

\end{document}